\documentclass[10pt,preprint]{elsarticle}
\usepackage{amssymb,amsmath,amscd}
\usepackage{amsfonts,latexsym,bm}
\usepackage[colorlinks=true,linkcolor=blue,citecolor=blue,urlcolor=blue]{hyperref}

\journal{Physics Letters B}

\def\be{\begin{equation}}
\def\ee{\end{equation}}
\def\bea{\begin{eqnarray}}
\def\eea{\end{eqnarray}}
\def\nn{\nonumber}
\def\p{\partial}

\begin{document}


\begin{frontmatter}

\title{Revisiting mass formulas of the four-dimensional Reissner-Nordstr\"{o}m-NUT-AdS
solutions in a different metric form}

\author{Di Wu and Shuang-Qing Wu\footnote{\textit{Email addresses}:
\texttt{wdcwnu@163.com} (Di Wu); \texttt{sqwu@cwnu.edu.cn} (Shuang-Qing Wu)}}


\address{School of Physics and Astronomy, China West Normal University, Nanchong,
Sichuan 637002, People's Republic of China}

\date{\today}

\begin{abstract}
Recently, the so-called ``consistent thermodynamics'' of the Lorentzian Reissner-Nordstr\"{o}m
(RN)-NUT-AdS$_4$ spacetimes has been pursued by a lot of efforts via different means. Among these
attempts, we had proposed a novel idea that ``The NUT charge is a thermodynamical multihair"
to successfully tackle with the subject. In this paper, we will adopt this strategy to reconsider
the mass formulas of the RN-NUT-AdS$_4$ solutions but written in an alternative form, which had
not been studied before in any existing literature and might be a most appropriate ansatz for
the higher dimensional multiply NUTty-charged AdS spacetimes without any constraint condition.
Here, we shall discuss the Christodoulou-Ruffini-like squared mass formula and
the first law as well as the Bekenstein-Smarr mass formula by introducing the secondary hair
$J_n = Mn$. For the sake of generality, we have introduced a dimensionless
constant $w$ into the constant factor $\Xi$ in the solution expression so that when $w = 1$,
all obtained results can reproduce those delivered in our previous work.
\end{abstract}


\end{frontmatter}

\section{Introduction}

In recent years, there is a great deal of interest in exploring consistent thermodynamics of the
Lorentzian Taub-NUT spacetimes in four dimensions \cite{PRD100-064055,JHEP0719119, CQG36-194001,
PLB798-134972,JHEP0520084,PRD100-104016,PLB832-137264,JHEP1022044,EPJC83-365,EPJC83-589,
PLB802-135270,IJMPD31-2250021,PRD103-024052,PRD106-024022,EPJP130-124,PTEP2020-043E05,PRD101-124011,
PRD105-124034,JHEP0321039,JHEP0821152,2208.05494,PRD100-101501,PRD105-124013} and
higher even dimensions \cite{2209.01757,2306.00062}. A most popular method advocated first in
\cite{CQG36-194001} to derive the mass formula relies on the Komar integral where the integration
path is split into different patches and consisted of two Misner string tubes, so that the contribution
of the Misner strings to the mass formulas is inlet via the ``$\Psi-\mathcal{N}$" (``Misner gravitational
charge") pair \cite{PRD100-064055,JHEP0719119,CQG36-194001,PLB798-134972,JHEP0520084,PRD100-104016,
PLB832-137264,JHEP1022044}. While this approach proves to be very effective in
the four-dimensional cases since the Komar mass at spatial infinity is finite, nevertheless, how
to extend it to the higher dimensional cases remains elusive up to date, simply
because the Komar integral at infinity is no longer finite in the case of higher even dimensional
NUT-charged spacetimes. On the other hand, thermodynamics of the four-dimensional Lorentzian Taub-NUT
spacetimes was analyzed \cite{PRD101-124011,PRD105-124034,JHEP0321039} on the basis of the traditional
thermodynamic method with which the Gibbs-Duhem relation was assured to be satisfied additionally.
Of course, a finite expression for the Gibbs (or Helmholtz) free energy must be achieved
from the beginning via the counter-term approach or the reference background method. Consequently,
this strategy might be a potential scheme that could be extended to investigate the higher even
dimensional NUT-charged cases. Besides, although the Iyer-Wald formalism was partially adopted in
\cite{JHEP0821152,2208.05494} to copy with the same topics in the four-dimensional NUT-charged
spacetimes, it is highly expected that the solution phase space method \cite{PRD93-044074,PRD104-044024}
could ultimately settle down the question and give a pleased final answer. However, nowadays there
is no related work and much has to be done along these two lines.

By contrast, a novel viewpoint that ``NUT charge as a multihair" was put forward in our previous
papers \cite{PRD100-101501,PRD105-124013}, where the Christodoulou-Ruffini-type squared mass formula
\cite{PRL25-1596,PRD4-3552} is derived from the beginning and acts as the starting point of the whole
work, then the first law and the Bekenstein-Smarr mass formula of almost all of the four-dimensional
dyonic NUT-charged spacetimes are consistently deduced by simply applying the ordinary Maxwell relation.
It is emphasized there that the NUT charge exhibits the two-fold facial features: rotation-like and
electric-charge-like characters. Quite recently, it is demonstrated \cite{2209.01757,2306.00062} that
our paradigm also works very well in the case of the higher dimensional static RN-Taub-NUT-AdS
spacetimes \cite{PLB632-537,PRD73-124039}.

Although we \cite{2209.01757,2306.00062} have taken the first step towards formulating the consistent
mass formulas of the higher even dimensional Lorentzian Taub-NUT spacetimes, the electrically neutral
solutions presented in Refs. \cite{PLB632-537,PRD73-124039} and studied there can be viewed essentially
as the special cases where all the NUT-charge parameters in the most generic multi-NUTty (AdS) solutions
\cite{PLB634-448,CQG21-2937} are set to equal. To resolve the constraint condition imposed on the
multi-NUTty AdS solutions, a general form of the solutions with the base space $\bigotimes^k{}S^2$
is presented for all higher even dimensions in the appendix \ref{AppA}. Since we are intending to
investigate the thermodynamic properties of these higher-dimensional Lorentzian Taub-NUT (AdS)
spacetimes with multiple NUT charge parameters in the future work, here as a warmup exercise we
shall first revisit the consistent thermodynamics of the RN-NUT-AdS$_4$ spacetime however expressed
in a different form. By the way, to the best of our knowledge, it is necessary to point out that
the solutions presented in this paper had never been studied before in any existing literature,
even for the four-dimensional cases, which deserves a detailed research in its own right.

The rest of this paper is organized as follows. In Sec. \ref{II}, we provide an alternative form
of the four-dimensional Lorentzian RN-Taub-NUT-AdS solution, and work out some thermodynamical
quantities that can be evaluated via the standard method. By introducing the
secondary hair $J_n = Mn$, we shall devote Sec. \ref{III} to discussing the consistent thermodynamics
by investigating the corresponding Christodoulou-Ruffini-like squared mass formula and the first
law as well as the Bekenstein-Smarr mass formula. Finally, we present our conclusions and give our
outlooks in Sec. \ref{IV}. In the appendix \ref{AppB}, we will establish the relation between the
usual form of the solution with $w=1$ and the alternative form when $w\not=1$. In
the appendix \ref{AppC}, we present a universal method to calculate the secondary hair $J_n$.

\section{Alternative form of the RN-NUT-AdS$_4$ solution
and some thermodynamical quantities}\label{II}

Distinct from its familiar expression of the RN-NUT-AdS$_4$ spacetime, here we present an
alternative form for the same solution, whose metric and Abelian gauge potential are
\bea
ds^2 &=& -\frac{f(r)}{r^2 +n^2}\Big(dt +\frac{2n\cos\theta d\phi}{\Xi} \Big)^2
 +\frac{r^2 +n^2}{f(r)}dr^2 \nn\\
&& +\frac{r^2 +n^2}{\Xi}\big(d\theta^2
 +\sin^2\theta d\phi^2\big) \, , \label{4dRNNUTAdS} \\
A &=& \frac{qr}{r^2 +n^2}\Big(dt +\frac{2n\cos\theta d\phi}{\Xi} \Big) \, , \label{4dRNNUTp}
\eea
where $\Xi = 1 +3(w-1)g^2n^2$, and the radial function reads
\be
f(r) = r^2 -2mr -n^2 +q^2 +g^2\big[r^4 +3(w +1)n^2r^2 -3wn^4\big] \, , \label{4dRNNUTf}
\ee
in which $q$, $n$ and $m$ are the electric charge, the NUT charge and the mass parameters,
respectively, $g$ is the gauge coupling constant, and $w$ is only a numerical
constant. When $w = 1$, the solution retains its familiar form that was adopted in our previous
work \cite{PRD100-101501}. If $w \ne 1$, then the angular two-dimensional spherical part of the
line element (\ref{4dRNNUTAdS}) is non-canonically normalized. The above general form of the
solution had not appeared before in any literature and thus had never been studied in any previous
work. In this paper, we are very interested in the case $w = 0$, whose higher even dimensional
analogues are presented in the appendix \ref{AppA}. But for the sake of generality,
hereafter we shall keep $w$ to be any arbitrary numerical constant, and shall not view it as a
thermodynamic variable in the below discussion. In the appendix \ref{AppB}, we will show explicitly
how the above alternative form when $w\not=1$ can be related to the usual RN-NUT-AdS$_4$ solution with
$w=1$ via some coordinate transformations and re-scaling of the solution parameters. Alternatively,
we will also present the second possible choice of the time coordinate and its corresponding solution.
Incidentally, we point out that the above general form of the NUT-charged AdS solution owns its
relatives in the gauged supergravity, for example, the simplest static dyonic NUT AdS solution in
the four-dimensional Kaluza-Klein supergravity theory must be given in a closely similar form. So
it is significant to study the above general form of the solution, since it not only acts as a
prototype of its higher even dimensional version, but also owns its relatives in the four-dimensional
gauged supergravity theories.

We begin by presenting some quantities that can be computed via the standard method. First, the
Bekenstein-Hawking entropy can be straightforwardly taken as one quarter of the horizon area:
\be
S = \frac{\mathcal{A}}{4} = \frac{\pi\big(r_+^2 +n^2\big)}{\Xi} \, ,
\ee
where is the location of the event horizon that is the largest root $r_+$ specified
by $f(r_+) = 0$. The Gibbons-Hawking temperature is directly proportional to the surface gravity
$\kappa$ on the event horizon:
\be
T = \frac{\kappa}{2\pi} = \frac{f^{\prime}(r_+)}{4\pi\big(r_+^2 +n^2\big)}
 = \frac{1 +3g^2\big(r^2 +wn^2\big)}{4\pi{}r_+}
 -\frac{q^2}{4\pi{}r_+\big(r_+^2 +n^2\big)} \, , \label{T}
\ee
in which a prime, here and hereafter, denotes the partial derivative with respective to its variable.

Second, as we did in our previous papers \cite{PRD100-101501,PRD105-124013}, we are only interested in
the global conserved electric charge that is measured at infinity and can be computed by using the
Gauss' law integral
\be
Q = \frac{-1}{4\pi}\int_{S^2_{\infty}}\star{} F = \frac{q}{\Xi} \, ,
\ee
and its corresponding electrostatic potential at the event horizon simply reads
\be
\Phi = (A_{\mu}\chi^{\mu})|_{r=r_+} = \frac{qr_+}{r_+^2 +n^2}\, ,
\ee
because a gauge has been chosen so that the gauge potential vanishes at infinity. In the above,
the timelike Killing vector $\chi = \p_t$ is normal to the event horizon, and the surface gravity
is computed via the definition: $\nabla^a(\chi^b\chi_b) = -2\kappa\chi^a$.

Third, one can adopt the conformal completion method \cite{PRD73-104036} or the Abbott-Deser method
\cite{NPB195-76} or the counter-term method \cite{PRD60-104001} to compute the
conserved mass $M$, and each technique gives the same result. Here we will follow the first one.
The idea is to perform a conformal transformation on the metric (\ref{4dRNNUTAdS}) to remove the
divergence in the integrals at the boundary (conformal infinity). After taking the $r\to \infty$
limit in the line element $ds^2/r^2$, one obtains the boundary metric
\be
ds_\infty^2 = -g^2\Big(dt -\frac{2n\cos\theta d\phi}{\Xi}\Big)^2
 +\frac{d\theta^2 +\sin^2\theta d\phi^2}{\Xi} \, .
\ee
Then the conserved charges $\mathcal{Q}[\xi]$ associated with the Killing vector $\xi$ can be
computed by
\be
\mathcal{Q}[\xi] = \frac{1}{8\pi{}g^3}\int_{S_\infty}dS_{\mu} \big(r\,
 C^{\mu}_{~\alpha\nu\beta}N^{\alpha} N^{\beta}\xi^{\nu}\big)_{r\to\infty} \, , \label{CCM}
\ee
where $C^{\mu}_{~\alpha\nu\beta}$ is the conformal Weyl curvature tensor, $N^\mu = \big[0,
-g^2r^2, 0, 0\big]$ is the vector normal to the boundary, and
\be
dS_t = g\sin\theta{} d\theta d\phi/\Xi
\ee
is the temporal component of the area vector in the three-dimensional conformal boundary. So the
conformal mass $M$ can be evaluated as
\be
M = \mathcal{Q}[\p_t] = \frac{m}{\Xi} \, .
\ee
Similarly, one can calculate the conformal dual mass by introducing the left-dual Weyl tensor as
\be
\widetilde{C}_{\mu\nu\rho\sigma} = \frac{1}{2}\epsilon_{\mu\nu\alpha\beta}
 C^{\alpha\beta}_{~~~\rho\sigma} \, ,
\ee
where $\epsilon_{\mu\nu\rho\sigma}$ is the Levi-Civita totally anti-symmetric tensor. Therefore,
by replacing the Weyl tensor $C^{\mu}_{~\alpha\nu\beta}$ in Eq. (\ref{CCM}) with the dual Weyl
tensor $\widetilde{C}^{\mu}_{~\alpha\nu\beta}$, the dual mass can be subsequently obtained as
follows:
\be
\widetilde{M} = \widetilde{\mathcal{Q}}[\p_t] = \frac{n\big[1 +(3w +1)g^2n^2\big]}{\Xi} \, .
\label{dualM}
\ee
Incidentally, here we would like to point out that the above conformal mass and its dual are
identical to the Yano-ADM charge and its dual charge \cite{JHEP0804045} associated with the
Killing-Yano tensor $k = db$ with $b = (r^2+n^2)(dt +2n\cos\theta{}d\phi/\Xi)$ and its Hodge
dual, respectively.

Finally, the NUT charge or gravitational magnetic (gravitomagnetic) charge can be computed as
\cite{PRD59-024009,CQG23-3951} (see the appendix \ref{AppC} for the detail),
\be\label{N}
N = \frac{n}{\Xi} \, .
\ee
Obviously, it is different from the above result (\ref{dualM}) of the dual mass.

\section{Mass formulas}\label{III}

In order to obtain the first law which is reasonable and consistent in both physical and
mathematical sense, we shall adopt the method used in Refs. \cite{PRD100-101501,PRD105-124013,
2209.01757,PLB608-251} to derive a meaningful Christodoulou-Ruffini-type squared mass formula.
First, substituting $r_+ = \sqrt{\Xi{}S/\pi -n^2}$ into the horizon equation: $4m^2r_+^2 =
\{r_+^2 -n^2 +q^2 +g^2[r_+^4 +3(w+1)n^2r_+^2 -3wn^4]\}^2$ yields the following identity:
\be
\frac{4m^2}{\Xi^2}\Big(\frac{S}{\pi\Xi} -\frac{n^2}{\Xi^2} \Big)
= \bigg\{\Big(\frac{S}{\pi\Xi} -\frac{2n^2}{\Xi^2}\Big)\big[1 +(3w +1)g^2n^2\big]
+\frac{g^2S^2}{\pi^2} +\frac{q^2}{\Xi^2}\bigg\}^2 \, . \label{sqm}
\ee
Inspired from this equation, it is clear that one can make the familiar substitutions
into the rhs of Eq. (\ref{sqm}): $m = M\Xi$, $n = N\Xi = 2N/\Gamma$, $q = Q\Xi$ and $g^2
= 8\pi{}P/3$, where $P = 3g^2/(8\pi)$ is the generalized pressure \cite{PRD84-024037},
and
\be
\Gamma = 1 +\sqrt{1 -12(w -1)g^2N^2} = 1 +\sqrt{1 -32(w -1)\pi{}N^2P} \, .
\label{Gamma}
\ee
According to the result presented in the appendix \ref{AppC}, we now further
introduce a secondary hair $J_n = Mn = mn/\Xi$, as we did before \cite{PRD100-101501,PRD105-124013}.
After finishing a little algebraic manipulation, one can arrive at an useful identity:
\bea
M^2 &=& \frac{\pi}{2S\Gamma}\bigg\{\bigg[1 +(3w +1)\frac{32\pi}{3\Gamma^2}N^2P \bigg]
\bigg(\frac{\Gamma}{2\pi}S -2N^2\bigg) +\frac{8P}{3\pi}S^2 +Q^2 \bigg\}^2 \nn \\
&& +\frac{\pi\Gamma}{2S}J_n^2 \, , \label{SQM}
\eea
which is our new Christodoulou-Ruffini-like squared mass formula for the four-dimensional
RN-NUT-AdS spacetime (\ref{4dRNNUTAdS}). We point out that Eq. (\ref{SQM}) when $w = 1$
consistently reduces to the one obtained in the familiar case of the RN-NUT-AdS$_4$ spacetime
\cite{PRD100-101501}.

At this step, we would like to give a new interpretation for Eq. (\ref{SQM}),
which is very crucial to our analysis done below. Otherwise, one may doubt that there exists
a mathematical inconsistency between the numbers of independent thermodynamical variables of
those of the free parameters appeared in the structure function $f(r)$ [Note that $w$ will not
be viewed as a solution parameter]. According the new viewpoint suggested in our recent papers
\cite{2209.01757,2306.00062}, equation (\ref{sqm}) can be thought of as representing a hypersurface
in the five-dimensional thermodynamical state space, the numbers of its variables ($m, n, S, q, g$)
exactly match with those of the solution parameters that appeared in the function $f(r)$. After
introducing an extra hair $J_n$, which is nothing but a kind of the higher-dimensional embedding
mapping, it becomes a hypersurface in the six-dimensional state space, as specified by Eq. (\ref{SQM}),
which now has six variables ($M, N, J_n, S, Q, P$). So, our below discussions will be based upon
this six-dimensional thermodynamical state space in which all its six variables could be regarded
as independent.

Now we are in a position to derive the differential and integral mass formulas
for the generic RN-NUT-AdS$_4$ spacetime (\ref{4dRNNUTAdS}). Since the secondary hair $J_n$ will
be treated as an independent variable, the above squared mass formula (\ref{SQM}) can be regarded
formally as a basic functional relation: $M = M(S, N, Q, P, J_n)$. As we did in Refs.
\cite{PRD100-101501,PRD105-124013,2209.01757,PRD103-044014,PRD101-024057,PRD102-044007},
differentiating it with respect to the thermodynamical variables $(S, N, Q, P, J_n)$ yields their
conjugate quantities, respectively, subsequently we can arrive at the differential and integral
mass formulas with their conjugate thermodynamic potentials given by the common Maxwell relations.

First, differentiation of the squared mass formula (\ref{SQM}) with respect to the entropy
$S$ leads to its conjugate Hawking temperature:
\bea
T &=& \frac{\p{}M}{\p{}S}\Big|_{(N,Q,P,J_n)}
 = \frac{1 +3g^2\big(r^2 +wn^2\big)}{4\pi{}r_+}
  -\frac{q^2}{4\pi{}r_+\big(r_+^2 +n^2\big)} \, ,
\eea
which is entirely identical to that given by Eq. (\ref{T}). Next, the electrostatic
potential $\Phi$ and the velocity-like potential $\omega_h$, which are conjugate to
$Q$ and $J_n$, respectively, are evaluated as
\bea
\Phi &=& \frac{\p{}M}{\p{}Q}\Big|_{(S,N,P,J_n)} = \frac{qr_+}{r_+^2 +n^2} \, , \\
\omega_h &=& \frac{\p{}M}{\p{}J_n}\Big|_{(S,N,Q,P)} = \frac{n}{r_+^2 +n^2} \, . \label{Phi}
\eea
Finally, the potential $\psi_h$ and the thermodynamical volume $V$, which are conjugate to
$N$ and $P$, respectively, can be computed as:
\bea
\psi_h &=& \frac{\p{}M}{\p{}N}\Big|_{(S,Q,P,J_n)} \nn \\
&=& \frac{n}{2(2 -\Xi)\big(r_+^2 +n^2\big)r_+}\bigg(4\big[2g^2\big(r_+^2 -3n^2\big)
 -1\big]r_+^2 +3(w -1)g^2 \nn \\
&& \times\Big\{r_+^4 -6n^2r_+^2 +n^4 +q^2\big(r_+^2 -n^2\big)
 +g^2\big[r_+^6 +(3w +2)n^2r_+^4 \nn \\
&& -3(2w +1)n^4r_+^2 +3wn^6 \big] \Big\}\bigg) \, , \label{psi_h} \\
V &=& \frac{\p{}M}{\p{}P}\Big|_{(S,N,Q,J_n)} \nn \\
&=& \frac{2\pi}{3\Xi(2 -\Xi)\big(r_+^2 +n^2\big)r_+}\bigg(2\big(r_+^4 +6n^2r_+^2
 -3n^4\big)r_+^2 +3(w -1)n^2 \nn \\
&& \times\Big\{r_+^4 -6n^2r_+^2 +n^4 +q^2\big(r_+^2 -n^2\big)
 +g^2\big[-r_+^6 +(3w -2)n^2r_+^4 \nn \\
&& -9(2w +1)n^4r_+^2 +3wn^6\big]\Big\} \bigg) \, . \label{volume}
\eea
It is obvious that only when $w = 1$, the electric charge parameter $q$ in the
expression of Eqs. (\ref{psi_h}) and (\ref{volume}) disappears, and the results reproduce the
familiar case in our previous work \cite{PRD100-101501}. When $w \ne 1$, the electric charge
parameter $q$ appears apparently in the above two expressions.

Now, one can check that both the first law and the Bekenstein-Smarr relation are completely
fulfilled
\bea
dM &=& TdS +\omega_h\, dJ_n +\psi_h\, dN +\Phi{}dQ +VdP \, , \quad \label{FL} \\
M &=& 2TS +2\omega_h\, J_n +\psi_h\, N +\Phi{}Q -2VP \, , \label{BS}
\eea
among all the aforementioned thermodynamical conjugate pairs.

Note that, when $w = 1$ and then $\Xi = 1$, the derived conjugate thermodynamic
volume
\be
V|_{w=1} = \frac{4\pi r_+\big(r_+^4 +6n^2r_+^2 -3n^4\big)}{3\big(r_+^2 +n^2\big)}
\label{Vw}
\ee
is not equal to $\widetilde{V} = 4\pi r_+(r_+^2 +3n^2)/(3\Xi)$ as can be computed via the definition
given in Ref. \cite{PRD84-024037}. If one prefers to match it to such a thermodynamic volume when
$w = 1$, then the dual (magnetic) mass (\ref{dualM}) can be further introduced as an additional
conserved charge into the first law and the Bekenstein-Smarr relation:
\bea
dM &=& TdS +\omega_h\, dJ_n +\widetilde{\psi_h}\, dN +\zeta{}d\widetilde{M}
 +\Phi{}dQ +\widetilde{V}dP \, , \quad \\
M &=& 2TS +2\omega_h\, J_n +\widetilde{\psi_h}\, N +\zeta\widetilde{M}
 +\Phi{}Q -2\widetilde{V}P \, ,
\eea
in which two new conjugate potentials are given by
\bea
\widetilde{\psi_h} &=& \psi_h -\frac{\zeta}{2 -\Xi}\big[1 +6(w +1)g^2n^2
 +3(w -1)(3w +1)g^4n^4\big] \, , \\
\zeta &=& \frac{r_+\big(r_+^2 -3n^2\big)}{4n\big(r_+^2 +n^2\big)}
 +\frac{3(w -1)}{4(3w +1)n\Xi\big(r_+^2 +n^2\big)r_+}\big[q^2\big(r_+^2 -n^2\big) \nn \\
&& -3n^2r_+^2 +n^4 +g^2\big(r_+^6 +5n^2r_+^4 -9wn^4r_+^2 +3wn^6\big) \big] \, .
\eea
As can be seen apparently from the above expression, it is not possible to reproduce the
thermodynamical volume $\widetilde{V}$ without the inclusion of the dual mass $\widetilde{M}$.

\section{Conclusions}\label{IV}

In this paper, we have employed our proposal that ``The NUT charge is a thermodynamical multihair"
to revisit the thermodynamics of the RN-NUT-AdS$_4$ spacetimes written in an unusual form. After
some thermodynamical quantities being evaluated via the standard method, we have
analyzed the consistent thermodynamics with the inclusion of a new secondary hair $J_n = Mn$. We
have obtained the consistent and reasonable Christodoulou-Ruffini-like squared mass formula, the
first law and the Bekenstein-Smarr mass formulas. When $w = 1$, they both reproduce
the previous results in the usual case \cite{PRD100-101501}.

Since the spacetime solution (\ref{4dRNNUTAdS}) is a special case for the generic even dimensional
multi-NUTty charged AdS solutions presented in the appendix \ref{AppA}, the present work acts as
a warmup excises for studying the high-dimensional Lorentzian Taub-NUT (AdS) spacetimes with multiply
NUT parameters \cite{PLB634-448,CQG21-2937} in the next step. On the other hand, we
have also adopted the ``$\Psi-\mathcal{N}$" formalism to investigate thermodynamics of the RN-NUT-AdS$_4$
solution given by the metric (\ref{4dRNNUTAdS}) and the gauge potential (\ref{4dRNNUTp}). It is
observed that one can derive a consistent Bekenstein-Smarr mass formula for arbitrary parameter $w$,
however, the first law is only consistent for the usual $w = 1$ case. This is a very odd result,
and we hope to report the related research soon.

\section*{Declaration of competing interest}
The authors declare that they have no known competing financial interests or personal
relationships that could have appeared to influence the work reported in this paper.

\section*{Data availability}
No data was used for the research described in the article.

\section*{Acknowledgments}
We are greatly indebted to the anonymous referee for his/her constructive comments
and suggestions to improve the presentation of this work. This work is supported by the National
Natural Science Foundation of China (NSFC) under Grants No. 12205243, No. 12375053, No. 11675130,
by the Sichuan Science and Technology Program under Grant No. 2023NSFSC1347, and by the Doctoral
Research Initiation Project of China West Normal University under Grant No. 21E028.

\appendix

\section{$D=2(k+1)$-dimensional multiply
NUTty-charged AdS solutions}\label{AppA}

The exact solutions of the higher-dimensional Lorentzian Taub-NUT (AdS) spacetimes with
multiple NUT charge parameters have already been given in Ref. \cite{PLB634-448,CQG21-2937},
but there exist additional constraint conditions that must be satisfied among in the solution
expressions. In this appendix, we present another general form of the ($2k+2$)-dimensional
Lorentzian Taub-NUT-AdS solutions with multiply NUT charge parameters as follows:
\be
ds^2 = -F(r)\Big(dt +2\sum_{i=1}^{k}\frac{n_i\cos\theta_i}{\Xi_i}d\phi_i\Big)^2
 +\frac{dr^2}{F(r)} +\sum_{i=1}^{k}\frac{r^2 +n_i^2}{\Xi_i}\big(d\theta_i^2
 +\sin^2\theta_i\, d\phi_i^2\big) \, ,
\ee
where $\Xi_i = \Xi -(2k +1)g^2n_i^2$, and
\bea
f(r) = \frac{r}{\prod\limits_{i=1}^k\big(r^2 +n_i^2\big)}\bigg\{\int\frac{\big[\Xi
 +(2k +1)g^2r^2\big]}{r^2}\prod\limits_{i=1}^k\big(r^2 +n_i^2\big)dr -2m\bigg\} \, ,
\eea
in which $n_i$ represents $k = D/2 -1$ distinct NUT charge parameters, and $\Xi$ is an arbitrary
constant that might be set to unity for convenience. It is worth mentioning that compared with the
solutions given in Refs. \cite{PLB634-448,CQG21-2937}, the above solutions have no need to introduce
any extra constraint. This formulation can be extended to the hyperbolic topological cases also,
but seems not suitable for the planar topological cases.

\section{Solution relation
 between $w=1$ and $w\not=1$}\label{AppB}

Denote with a bar the case with $w=1$, the usual standard form of the four-dimensional
Reissner-Nordstr\"{o}m NUT-charged AdS solution reads
\bea
d\bar{s}^2 &=& -\frac{\bar{f}(\bar{r})}{\bar{r}^2 +\bar{N}^2}(d\bar{t}
 +2\bar{N}\cos\theta{} d\phi)^2 +\frac{\bar{r}^2 +\bar{N}^2}{\bar{f}(\bar{r})}d\bar{r}^2 \nn \\
&& +\big(\bar{r}^2 +\bar{N}^2\big)\big(d\theta^2 +\sin^2\theta{} d\phi^2\big) \, , \\
\bar{A} &=& \frac{\bar{Q}\bar{r}}{\bar{r}^2 +\bar{N}^2}(d\bar{t}
 +2\bar{N}\cos\theta{} d\phi) \, ,
\eea
where the radial function is
\be
\bar{f}(\bar{r}) = \bar{r}^2 -2\bar{M}\bar{r} -\bar{N}^2 +\bar{Q}^2
 +g^2\big(\bar{r}^4 +6\bar{N}^2\bar{r}^2 -3\bar{N}^4\big) \, .
\ee

After performing the following replacements:
\be
\bar{r} = \frac{r}{\sqrt{\Xi}} \, , \quad \bar{N} = \frac{n}{\sqrt{\Xi}} \, , \quad
\bar{M} = \frac{m}{\Xi^{3/2}} \, , \quad \bar{Q} = \frac{q}{\Xi} \, ,
\ee
the above metric and the gauge potential become
\bea
d\hat{s}^2 &=& -\frac{f(r)}{\Xi\big(r^2+n^2\big)}\Big(d\bar{t}
 +\frac{2n\cos\theta}{\sqrt{\Xi}} d\phi\Big)^2 +\frac{r^2+n^2}{f(r)}dr^2 \nn \\
&& +\frac{r^2 +n^2}{\Xi}\big(d\theta^2 +\sin^2\theta{} d\phi^2\big) \, , \label{le1} \\
\hat{A} &=& \frac{qr}{\sqrt{\Xi}\big(r^2 +n^2\big)}\Big(d\bar{t}
 +\frac{2n\cos\theta}{\sqrt{\Xi}} d\phi\Big) \, . \label{gp1}
\eea
where the structure function is: $f(r) = \Xi^2\bar{f}(\bar{r}) = (r^2 -n^2)\Xi -2mr +q^2
 +g^2\big(r^4 +6n^2r^2 -3n^4\big)$, and $\Xi = 1 +3(w-1)g^2n^2$.

Then, it is not difficult to find that the new form of the metric (\ref{4dRNNUTAdS}) and
the gauge potential (\ref{4dRNNUTp}) with the radial function (\ref{4dRNNUTf}) can be
obtained by further setting the time coordinate: $t = \bar{t}/\sqrt{\Xi}$. Alternatively,
it is also possible to adopt another time coordinate: $\tilde{t} = \sqrt{\Xi}\,\bar{t} =
\Xi\, t$, so that the solution is written as
\bea
d\tilde{s}^2 &=& -\frac{f(r)}{\Xi^2\big(r^2+n^2\big)}(d\tilde{t} +2n\cos\theta{} d\phi)^2
 +\frac{r^2+n^2}{f(r)}dr^2 \nn \\
&& +\frac{r^2 +n^2}{\Xi}\big(d\theta^2  +\sin^2\theta{} d\phi^2\big) \, , \label{le2} \\
\tilde{A} &=& \frac{qr}{\Xi\big(r^2 +n^2\big)}(d\tilde{t} +2n\cos\theta{} d\phi) \, .
\label{gp2}
\eea
This latter form can also be extended to higher even dimension AdS solutions with equal
NUT-charges, but it seems inappropriate for making a generalization to the multiply NUTty
cases.

There is another possibility to make a rescaling of the angular coordinate $\phi
= \Xi\tilde{\phi}/K$ in the metric (\ref{4dRNNUTAdS}) and the gauge potential (\ref{4dRNNUTp}),
so that the resulting solution represents the four-dimensional NUT-AdS spacetimes pierced by a
cosmic string.

Anyway, the above arguments reveal that there exists a subtle issue to choose the timelike
Killing vector in the four-dimensional NUT-charged AdS case, and we will abandon to discuss
the solution given by Eqs. (\ref{le2}) and (\ref{gp2}) in the main context.

\section{A universal method to calculate the secondary hair $J_n$}\label{AppC}

The purpose of this appendix is to provide a universal definition to evaluate the
secondary hair $J_n$, which is borrowed from the appendix of Ref. \cite{2306.00062}. First, we would
like to point out that the mass $M = m/\Xi$ can be computed from the usual Komar integral with a
massless background subtraction. Correspondingly, one can also use this background subtracted
Komar integral to get the secondary hair $J_n$, which is given by \cite{2306.00062}:
\be
J_n = \frac{-1}{16\pi}\int_{S^2}r(\Xi -\bar{\Xi}) = \frac{mn}{2\Xi}\int_0^{\pi} \sin\theta d\theta
 = \frac{mn}{\Xi} = Mn \, , \label{Jne}
\ee
where the two-form $\Xi = d\xi$ is the usual Komar superpotential written in the language of
differential forms as follows:
\be
\Xi = -\p_r{}f(r){} dr\wedge \Big(dt +\frac{2n\cos\theta}{\Xi}{}d\phi\Big)
 +2nf(r)\frac{\sin\theta}{\Xi}{}d\theta \wedge{} d\phi \, ,
\ee
which is associated with the timelike vector $\xi^a = (\p_t)^a$, whose dual one-form is: $\xi
= -f(r)\big(dt +2n\cos\theta{}d\phi/\Xi\big)$. The related background 1-form is: $\bar{\xi} =
-\bar{f}(r)\big(dt +2n\cos\theta{}d\phi/\Xi\big)$, in which $\bar{f}(r) = f(r)|_{m=0}$. Apart
from the need of the massless background substraction, the linear factor $r$ in the definition
(\ref{Jne}) clearly demonstrates that the secondary hair $J_n$ is a sub-leading dual charge.

Likewise in the Ref. \cite{PRD59-024009}, the NUT charge is defined as:
\be
N = \frac{-1}{8\pi}\int_{S^2}d\hat{\xi} = \frac{n}{2\Xi}\int_0^{\pi} \sin\theta d\theta
 = \frac{n}{\Xi} \, ,
\ee
where $\hat{\xi} = \xi/g_{tt} = dt +2n\cos\theta{}d\phi/\Xi$, and the result is presented
in Eq. (\ref{N}).

By the way, we would like to mention that as already shown in Ref. \cite{2306.00062}
we need not to in the secondary $Q_n \simeq{} qn$ since we only focus on the electrically
charged case in the main context.

\end{document}